\documentclass[reprint,superscriptaddress,amsmath,amssymb,aps,onecolumn,]{revtex4-2}

\usepackage{graphicx}
\usepackage{bm}
\usepackage[caption=false,font=normalsize]{subfig}
\usepackage{braket}
\usepackage{qcircuit}
\usepackage{setspace}

\makeatletter
\newcommand\thefontsize{The current font size is: \f@size pt}
\def\blfootnote{\xdef\@thefnmark{}\@footnotetext}

\usepackage{microtype}

\usepackage{xcolor}
\definecolor{lightblue}{rgb}{0.63, 0.74, 0.78}
\definecolor{seagreen}{rgb}{0.18, 0.42, 0.41}
\definecolor{orange}{rgb}{0.85, 0.55, 0.13}
\definecolor{silver}{rgb}{0.69, 0.67, 0.66}
\definecolor{rust}{rgb}{0.72, 0.26, 0.06}

\colorlet{lightsilver}{silver!30!white}
\colorlet{darkorange}{orange!75!black}
\colorlet{darksilver}{silver!65!black}
\colorlet{darklightblue}{lightblue!65!black}
\colorlet{darkrust}{rust!85!black}

\usepackage[colorlinks=true,linkcolor=darkrust,citecolor=darklightblue,urlcolor=darksilver]{hyperref}

\usepackage{cleveref}

\usepackage[margin=1in]{geometry}

\usepackage{bm}
\newcommand{\ve}[1]{\bm{#1}}

\newcommand{\br}{\ve{r}}

\newcommand{\bc}{\ve{c}}

\makeatletter
\def\Dated@name{}
\makeatother

\begin{document}

\title{Fully quantum algorithm for mesoscale fluid simulations \\  with application to partial differential equations}

\author{Sriharsha Kocherla}
\affiliation{School of Computational Science \& Engineering, Georgia Institute of Technology, Atlanta, GA}

\author{Zhixin Song}
\affiliation{School of Physics, Georgia Institute of Technology, Atlanta, GA}

\author{Fatima Ezahra Chrit}
\affiliation{George W.\ Woodruff School of Mechanical Engineering, Georgia Institute of Technology, Atlanta, GA}
\affiliation{School of Computational Science \& Engineering, Georgia Institute of Technology, Atlanta, GA}

\author{Bryan Gard}
\affiliation{CIPHER, Georgia Tech Research Institute, Atlanta, GA}

\author{Eugene~F.~Dumitrescu}
\affiliation{Oak Ridge National Laboratory, Oak Ridge, TN}

\author{Alexander Alexeev}
\affiliation{George W.\ Woodruff School of Mechanical Engineering, Georgia Institute of Technology, Atlanta, GA}

\author{Spencer H. Bryngelson}
\affiliation{School of Computational Science \& Engineering, Georgia Institute of Technology, Atlanta, GA}
\affiliation{George W.\ Woodruff School of Mechanical Engineering, Georgia Institute of Technology, Atlanta, GA}
\affiliation{Daniel Guggenheim School of Aerospace Engineering, Georgia Institute of Technology, Atlanta, GA}

\date{}

\begin{abstract}
    \begin{center}
            \textbf{Abstract} 
    \end{center}
    Fluid flow simulations marshal our most powerful computational resources.
    In many cases, even this is not enough.
    Quantum computers provide an opportunity to speed up traditional algorithms for flow simulations.
    We show that lattice-based mesoscale numerical methods can be executed as efficient quantum algorithms due to their statistical features.
    This approach revises a quantum algorithm for lattice gas automata to reduce classical computations and state preparation at every time step.
    For this, the algorithm approximates the qubit relative phases and subtracts them at the end of each time step.
    Phases are evaluated using the iterative phase estimation algorithm and subtracted using single-qubit rotation phase gates.
    This method optimizes the quantum resource required and makes it more appropriate for near-term quantum hardware.
    We also demonstrate how the checkerboard deficiency that the D1Q2 scheme presents can be resolved using the D1Q3 scheme.
    The algorithm is validated by simulating two canonical PDEs: the diffusion and Burgers' equations on different quantum simulators. 
    We find good agreement between quantum simulations and classical solutions for the presented algorithm.
\end{abstract}


\blfootnote{Corresponding author: \url{shb@gatech.edu} \\[-5pt]

\noindent This manuscript has been authored by UT-Battelle, LLC, under Contract No.~DE-AC0500OR22725 with the U.S.~Department of Energy. The United States Government retains and the publisher, by accepting the article for publication, acknowledges that the United States Government retains a non-exclusive, paid-up, irrevocable, world-wide license to publish or reproduce the published form of this manuscript, or allow others to do so, for the United States Government purposes. The Department of Energy will provide public access to these results of federally sponsored research in accordance with the DOE Public Access Plan.}

\maketitle

\section{Introduction}\label{s:introduction}

Solving partial differential equations (PDEs) is necessary for modeling most scientific problems. 
However, it is computationally expensive, especially for large domains and high-dimensional configuration spaces. 
For example, in computational fluid dynamics, the cost of performing a direct numerical simulation of turbulence increases with Reynolds number, representing the ratio of inertial to viscous effects in the flow, as $\mathcal{O}(\mathrm{Re}^3)$~\citep{pope2000turbulent}. 
Such high computational expense motivates a broad category of work to achieve algorithmic speedups for PDE solvers.

Quantum computers are emerging as an increasingly viable tool for algorithm speedup enabled by the principles of quantum mechanics, such as superposition and entanglement.
In some cases, exponential speedups over classical computer architectures (like that of von~Neumann) are possible~\citep{grover1996fast, shor1994algorithms}.
In the sciences, algorithmic complexity improvements are seen in quantum chemistry~\citep{cao2019quantum}, machine learning~\citep{wittek2014quantum, schuld2014quest, marquardt2021machine}, finance~\citep{egger2020quantum, bouland2020prospects}, and more.

Quantum algorithms for solving PDEs have also gained attention. 
\citet{harrowQuantumAlgorithmSolving2009} demonstrated the first quantum algorithm, HHL, for solving linear systems of equations with exponential speedup.
Such improvements can be marshaled in the PDE space for solving discretized systems of equations via, for example, the method of lines.
\citet{childs2020quantum} proposed a quantum algorithm for linear ordinary differential equations (ODEs) based on spectral methods, providing a global approximation to the solution using linear combinations of basis functions.
\citet{childsHighprecisionQuantumAlgorithms2021} improved the complexity of quantum algorithms for linear PDEs using adaptive-order finite difference and spectral methods.
Other approaches to solving PDEs were proposed by \citet{berry2014high}; for example, reducing an ODE system via discretization and solving it via the appropriate quantum algorithms. 
\citet{gaitanFindingFlowsNavier2020} used this approach to solve the 1D Navier--Stokes equations for the flow through a convergent-divergent nozzle.
However, these methods require deep circuits and many quantum gates to achieve results comparable to what one can solve on even a modern laptop.
This limitation is prohibitive for most of the algorithms mentioned above but paints a bright future for larger, fault-tolerant quantum devices that can better reject gate noise.

In the near term, variational quantum algorithms are well-suited for current NISQ (noisy intermediate-scale quantum) devices.
For example, hybrid quantum--classical procedures can evaluate the solution quality via a cost function using a quantum computer and optimize variational parameters using a classical computer. 
Thus, variations methods enable quantum algorithms with relatively shallow gate depths and qubit counts.
\citet{bravo-prieto_variational_2020} introduced the variational quantum linear solver (VQLS), which uses the Hadamard test to solve a linear system.
\citet{liu_variational_2021} showed that a Poisson problem can be solved by first converting it to a linear system amenable to the Quantum Alternating Operator Ansatz (QAOA) algorithm.
However, variation algorithms require a (sometimes) unclear ansatz and do not guarantee convergence or speedup.
The continued tension between these quantum algorithm threads has contributed to improving each.

Because of the linearity of quantum mechanics, quantum algorithms aiming to solve nonlinear PDEs are immature compared to the linear case. 
\citet{liuEfficientQuantumAlgorithm2021} proposed the first quantum algorithm for dissipative nonlinear differential equations using the method of Carleman linearization, which maps the system of nonlinear equations to an infinite-dimensional system of linear differential equations, which are then solved using quantum linear system algorithms. 
Their algorithm was tested to solve the 1D Burgers' equation.
\citet{kyriienko_solving_2021} used the variational method based on differentiable quantum circuits to solve nonlinear differential equations.
Their method solved a quasi-1D approximation of Navier--Stokes equations.
\citet{lubasch2020variational} used multiple copies of variational quantum states and tensor networks to solve nonlinear PDEs. 

Other promising numerical methods include mesoscale strategies, which operate between molecular and continuum scales.
Lattice methods are a common example and are well suited for quantum computation because they are intrinsically statistical, resolving only samples or probabilities of the fictitious particles they comprise. 
Moreover, they are based on simple mathematical calculations and are suitable for parallel computation because interactions between lattice nodes are linear, and non-linearity enters during a local collision step. 
The lattice Boltzmann method (LBM) is a common approach that solves the Boltzmann transport equation. 
\citet{mezzacapo2015quantum} developed the first quantum simulator following a lattice kinetic formalism to solve fluid dynamics transport phenomena. 
\citet{todorova2020quantum} solved the collisionless Boltzmann equation on a quantum computer, inspired by quantum algorithms solving the Dirac equation. 
\citet{budinskiQuantumAlgorithmAdvection2021} proposed a novel quantum algorithm that solves the advection--diffusion equation by the LBM and extended it to solve the 2D Navier--Stokes equations using stream function--vorticity formulation~\citep{budinskiQuantumAlgorithmNavierStokes2021}. This technique has a logarithmic scaling of qubits relative to lattice grid size.
\citet{itani2022analysis} explored the Carleman linearization of the collision term of the lattice Boltzmann equation. 
Using this linearization technique, \citet{itani2023quantum} proposed a quantum lattice Boltzmann algorithm where collision and streaming are achieved by unitary evolution.
\citet{schalkers2024importance} proposes a space-time encoding method to implement unitary collision and streaming operators without linearization.

The LBM originates from the lattice gas automata (LGA), a cellular automaton that can simulate fluid flows. 
A quantum computer can be viewed as a quantum cellular automaton (QCA), where each cell is a quantum system with a state depending on neighboring cells.
Quantum LGA is a subclass of QCA that represents particles interacting under physical constraints and evolving on a lattice. 
The main drawback of the LGA is statistical noise, though this vanishes when crafted for mesoscopic scales.
\citet{yepezQuantumLatticegasModel2001} introduced the first quantum LGA (QLGA) algorithm for simulating fluid flow at the mesoscopic scale using a discretized Boltzmann transport equation. 

The original algorithm was designed for a specialized Type-II quantum computer, proposed in the early 2000s~\citep{yepez2001type}. 
Such computers consisted of several small quantum computers connected by classical communication channels, allowing qubit shifts during the algorithm's streaming step~\citep{yepez2001type}.
Type~II architectures no longer exist.
Instead, we present an algorithm more suitable for current architectures that avoids some assumed-efficient operations on previous Type~II devices.
For example, many QLGA/LBM algorithms require estimating the distribution functions needed to re-prepare the quantum states at each time step.
This process can be prohibitively expensive on current, and likely future, quantum devices.

Herein, we present an approach that overcomes the limitations of previous lattice-based quantum algorithms shown by Yepez~\citep{yepez2001type, yepez2002quantum}.
We focus on improving the efficiency and scalability of previous QLGA techniques.
The cost of general quantum state preparation scales exponentially with respect to the qubit count~\citep{nielsenQuantumComputationQuantum2010}.
This expensive step, assumed in previous quantum lattice-based methods using amplitude encoding, presents scaling issues for utility-scale simulations.
Our work seeks to mitigate these expense and scalability issues.
The proposed algorithm uses a quantum implementation of the streaming step, approximates the qubit relative phases, and subtracts them at the end of each time step, avoiding repeated encoding operations and delaying the need for measurement to the end of the computation. 

We do not interrogate the practicality of the presented algorithmic strategies on actual quantum devices but rather on simulators.
The algorithm presented herein is not appropriate for solving practical fluid dynamics problems, in full, on current quantum devices, for which there is perhaps no known algorithm at the time of writing.
Instead, simulators help achieve validation and facilitate analysis.
Specifically, we implement our algorithm in Qiskit~\cite{aleksandrowicz2019qiskit} and XACC for the diffusion and Burgers' equations.

This manuscript continues as follows.
\Cref{Sec2} describes the classical lattice gas automata and lattice Boltzmann method. 
\Cref{Sec3} introduces the quantum counterpart of the classical LGA. 
\Cref{Sec4} presents the QLGA variants we propose, including quantum streaming, the D1Q3 scheme, and a restless QLBM algorithm. 
We validate these variants by testing them to solve the 1D diffusion and Burgers' equations. 
Last, \cref{Sec5} compares different quantum simulation strategies for the proposed algorithm using Qiskit and XACC.

\section{Classical lattice methods} \label{Sec2}

The lattice gas automata (LGA) method is a cellular automaton that simulates fluids~\citep{frisch2019lattice, frisch2019lattice2}. 
The LGA state is described by \textit{occupation numbers}: Boolean variables indicating whether or not a fictitious particle is present at a specific lattice node, moving in a certain lattice direction at a specific time. 
These fictitious particles obey local collision rules that conserve mass and momentum. 

\begin{figure}
    \includegraphics[trim={0 0.6cm 0 0.3cm},clip]{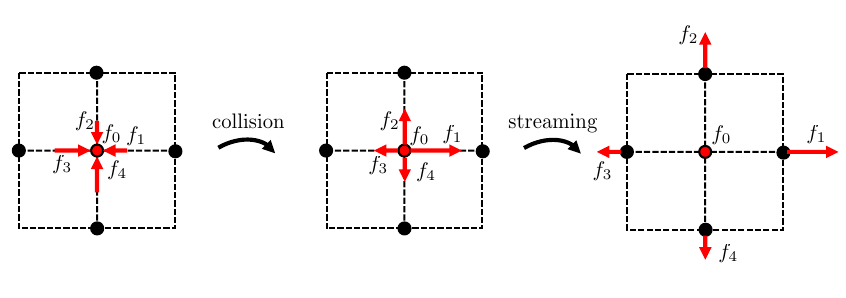}
    \caption{Illustration of collision and streaming steps for a single time-step of a D2Q5 scheme.}
    \label{fig:lbm} 
\end{figure}

The main drawback of LGA is noise, which can be reduced via coarse-graining over large domains or time intervals but remains costly in computing time and memory. 
Instead, LGA can be directly modeled at the mesoscopic scale with the Boltzmann equation~\citep{mcnamara1988use, wolf2004lattice}. 
This approach is called the lattice Boltzmann method (LBM), which eliminates the LGA noise by replacing Boolean variables with continuous distributions, i.e., the mean occupation numbers, also called density distribution functions. 
We take this approach herein, which is thus a QLBM-type technique.
The LBM approach replaces the collision rule with a continuous function called the collision operator.
The distribution functions $f_{\alpha}$ ($0 \leq f_{\alpha} \leq 1$) evolve in time according to the kinetic equation 
\begin{gather} \label{eq1}
    f_{\alpha}(\br+ \bc_{\alpha} \Delta t, t+\Delta t) = f_{\alpha}(\br, t) + \Omega_{\alpha}[f_{\alpha}(\br, t)],
\end{gather}
where $\br$ is the position of a lattice site, $\bc_{\alpha}$ is the lattice velocity, $\Delta t$ is the time step, $t$ is time, $\Omega_{\alpha}$ is the collision operator, and ${\alpha}$ refers to the velocity discretization index.
During one time step, as illustrated in \cref{fig:lbm}, the computation evolves through two processes: collision, where particles meet at the same lattice site, and their distribution functions are evaluated according to the collision operator $\Omega_{\alpha}$, and streaming, where particles shift to the neighboring sites following their lattice directions. 
Collision causes the relaxation of all local distribution functions to an equilibrium where $\Omega_{\alpha}[f_{\alpha}=f_{\alpha}^\mathrm{eq}]=0$.

Each discretization of the velocity space is described by its number of spatial dimensions $a$ and velocity directions $b$, with scheme notation D$a$Q$b$. 
For example, a D1Q2 scheme uses a 1D lattice with two velocity directions, i.e., a right-going particle and a left-going particle for each lattice site.

Macroscopic variables, like the fluid density or velocity, are determined from the moments of the distribution functions~\citep{chen1998lattice}.
For example, the density at lattice site $i$ corresponds to
\begin{gather}
    \rho_i = \sum_{\alpha=1}^b f_{\alpha,i}.
    \label{e:density}
\end{gather}
Herein, we divide 1D spatial domains into $N_s$ uniformly spaced lattice sites $ i=0, \dots, N_s-1$.
For example, domain $X \in [0,L]$ has site locations $x_i = i \Delta x$ where $\Delta x = L/N_s$ is the lattice spacing, usually set to $1$.

\section{Quantum lattice methods} \label{Sec3}

\begin{figure}[h]
    \includegraphics[width=0.5\textwidth]{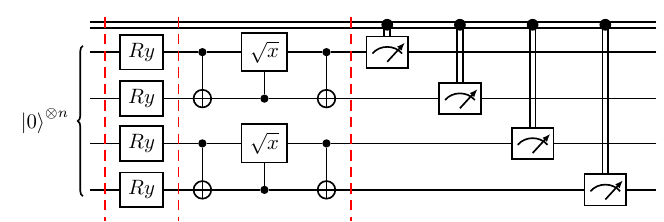}
    \caption{QLGA with classical streaming~\citep{yepezQuantumLatticegasModel2001}. $R_y$ is a rotation gate by the angle $\theta_{\alpha} = 2 \arccos{\sqrt{1-f_{\alpha}(x_i, t)}}$ for the qubit of state $\ket{q_{\alpha}(x_i, t)}$. 
    The controlled $\sqrt{x}$ gate can be decomposed efficiently on the latest IBM Heron and Eagle processors, which consists of a two-qubit controlled phase rotation $P(\pi/2)$ gate between two Hadamard gates on the target qubit.
    }
    \label{fig:qlga} 
\end{figure}

The quantum LGA (QLGA) is a measurement-based algorithm that introduces a superposition of qubit states within a small spatial region and for a short period. 
It directly maps the bits used in a classical lattice gas to qubits and models the lattice gas at the mesoscopic scale, which can be described using the lattice Boltzmann equation.
\Cref{fig:qlga} shows an example of the associated quantum circuit used to solve the 1D diffusion equation using the D1Q2 scheme defined in the previous section. 
The algorithm has four operations: (re)initializing the qubit states via \textit{encoding}, \textit{collision} at each lattice site, measuring the qubit occupancy via \textit{localization}, and \textit{streaming} the qubits on the lattice appropriately.

Encoding and collision are performed in the quantum space, but the measurement step and classical streaming computation mean the quantum state must be re-encoded. 
The encoding step in QLGA \citep{yepezQuantumLatticegasModel2001} scales linearly in the number of qubits required for encoding with respect to the size of the simulation.
There are QLBM methods \citep{budinskiQuantumAlgorithmAdvection2021, budinskiQuantumAlgorithmNavierStokes2021} that scale logarithmically with respect to the size of the grid.
However, these entail a prohibitively large number of quantum gates (in many cases more than $10^5$ for even a small, simplified problem) a trade-off that may be unappealing.

The system wavefunction can be written as
\begin{gather} \label{eq2}
\ket{\Psi(x_0, x_1, \dots, x_{N_s-1}, t)} = \bigotimes_{i=0}^{N_s-1} \ket{\Psi(x_i, t)}.
\end{gather}
Following Dirac's bra-ket notation, the ket vector $\ket{\Psi(x_i, t)}$ represents the on-site state, i.e., at the lattice site of coordinate $x_i$, of dimension $2^b$. 
In the D1Q2 scheme, each lattice site comprises two particles, one moving to the right and the other to the left. 
The first step of the QLGA algorithm is mapping each particle to one qubit by encoding the occupancy probability $f_{\alpha}(x_i, t)$ to $\alpha$-th qubit as
\begin{gather} \label{eq3}
    \ket{q_{\alpha}(x_i, t)} = 
        \sqrt{1-f_{\alpha}(x_i, t)} \ket{0} + \sqrt{f_{\alpha}(x_i, t)} \ket{1} 
    \quad \mathrm{for} \quad 
    i = 0, \dots, N_s-1 
    \quad \mathrm{and}  \quad 
    \alpha = 1,2. 
\end{gather}
This step is implemented via the $R_y(\theta_{\alpha})$ rotation gate with angle $\theta_{\alpha} = 2 \arccos{\sqrt{1-f_{\alpha}(x_i, t)}}$ for the qubit of state $\ket{q_{\alpha}(x_i, t)}$.
Then, the full quantum state at each lattice site $x_i$ is the tensor product state of two qubits
\begin{gather}\label{eq4}
    \ket{\Psi(x_i, t)} = \ket{q_1(x_i, t)} \otimes \ket{q_2(x_i, t)} \quad \mathrm{for} \quad i=0, \dots, N_s-1.
\end{gather} 
The relative phase of the qubits is set to zero in this step.
Then, QLGA applies the collision locally to each lattice site, a unitary operation that relaxes the distribution functions to equilibrium. 
The post-collision state is
\begin{gather} \label{eq5}
\ket{\Psi'(x_i, t)} = U \ket{\Psi(x_i, t)} 
\end{gather}
where $U$ is the collision operator that causes local quantum entanglement of the on-site qubits. 
It is implemented via a unitary gate with a block-diagonal  entangling $U(2)$ matrix:
\begin{gather} \label{eq6}
\begin{pmatrix}
1 & 0 & 0 & 0\\
0 & e^{i\phi} e^{i\xi} \cos{\theta} & e^{i\phi} e^{i\zeta} \sin{\theta} & 0\\
0 & -e^{i\phi} e^{-i\zeta} \sin{\theta} & e^{i\phi} e^{-i\xi} \cos{\theta} & 0\\
0 & 0 & 0 & 1\
\end{pmatrix}
\end{gather}
where $\phi$, $\xi$, $\zeta$, and $\theta$ are Euler angles chosen to satisfy the PDE of interest in the continuum limit.
We choose $\phi=-\pi/4$, $\xi=0$, $\zeta=\pi/2$ and $\theta=\pi/4$ to model the diffusion equation and $\phi=0$, $\xi=\zeta$ and $\theta=\pi/4$ to model Burgers' equation~\citep{yepezQuantumLatticegasModel2001a, yepez2002quantum}.

The next step is measuring all the $2N_s$ qubits associated with $N_s$ lattice sites. 
Notice the local nature of these measurements.
The algorithm starts with a product state.
The encoding (re-preparation) and collision steps are local operations inside each lattice site and, hence, do not change the structure of the product state.
Thus, the number of measurements required in each time step scales linearly with the number of lattice sites.
Then, the post-collision occupancy probabilities $f'_{\alpha}(x_i, t)$ are computed from measurement outcomes. 
The non-unitary measurement operation destroys the quantum superposition and entanglement that the collision step may have caused. 
The post-collision occupancy probability is
\begin{gather} \label{eq7}
    f'_{\alpha}(x_i, t) = \bra{\Psi'(x_i, t)}n_{\alpha}\ket{\Psi'(x_i, t)}
\end{gather}
where $n_\alpha$ are the number operators (observables that can be measured to count the number of particles) for $\alpha=1,2$.

Streaming is performed by shifting the qubits to their neighboring positions according to their right ($\alpha=1$) or left ($\alpha=2$) lattice directions. 
The post-streaming occupancy probabilities can be written as
\begin{gather} \label{eq8}
    f_{\alpha}(x_i, t+\Delta t) = f'_{\alpha}(x_i+e_{\alpha}\Delta x, t), 
\end{gather}
where $e_1=-1$ and $e_2=1$ are the qubits directions, $\Delta x$ is the spacing, and $\Delta t$ is the time step.

These four operations can be encapsulated in the following equation:
\begin{gather} \label{eq9}
    f_{\alpha}(x_i +e_{\alpha}\Delta x, t+\Delta t) = f_{\alpha}(x_i, t) + \Omega_{\alpha}[f_{\alpha}(x_i, t)]
\end{gather}
which is in the form of the classical lattice Boltzmann equation with the collision operator given by
\begin{gather} \label{eq10}
   \Omega_{\alpha}[f_{\alpha}(x_i, t)] = \bra{\Psi(x_i, t)}( U^{\dagger} n_{\alpha} U - n_{\alpha} )\ket{\Psi(x_i, t)}.
\end{gather}
Equilibrium distributions can be found by satisfying the equilibrium condition:   $\Omega_{\alpha}[f_{\alpha}=f_{\alpha}^\mathrm{eq}]=0$. 
For the diffusion equation, we get equal equilibrium probabilities $f_{\alpha}^\mathrm{eq}=\rho/2$ for $\alpha=1,2$~\citep{yepezQuantumLatticegasModel2001a}. 
For Burgers' equation, the equilibrium distribution corresponds to $f_{\alpha}^\mathrm{eq}=\rho/2  + e_{\alpha}/2(\sqrt{2} - \sqrt{1+(\rho-1)^2})$~\citep{yepezQuantumLatticegasModel2002}.
The equilibrium distributions for diffusion and Burgers' equations differ due to the inherent differences in the underlying physical processes.
The equilibrium state of the diffusion equation is characterized by maximum entropy and minimum energy.
This equation results in a smooth, uniform distribution.
Meanwhile, the equilibrium state of Burgers' equation is characterized by the balance of convective and diffusive forces.

This streaming step was designed to run classically on Type-II quantum computers~\citep{yepez2001type}. 
One time step of QLGA consists of the four operations described above. 
Unlike most quantum gate models, where measurement is performed at the end of the computation, QLGA requires measurement at each time step to extract the distribution functions, which are then used to encode the next step. 
Next, we present a method for improving these unnecessary computations that is more appropriate for modern quantum hardware.

\section{A fully-quantum lattice Boltzmann algorithm} \label{Sec4}

This section presents a single-quantum computer version of QLGA that avoids the Type-II quantum computers of previous studies.
We formulate a fully-quantum lattice Boltzmann method algorithm (QLBM) and verify it against known, classical solutions.
Since our techniques solve the lattice Boltzmann equation at the mesoscopic scale using density distributions instead of boolean operators, we denote our QLGA variants as QLBM.

\subsection{Quantum streaming}

\begin{figure}[t]
    \centering
    \begin{tabular}{c c}
        \multicolumn{1}{l}{\footnotesize (a)} & 
        \multicolumn{1}{l}{\footnotesize (b)} \\
        \includegraphics[width=0.5\textwidth]{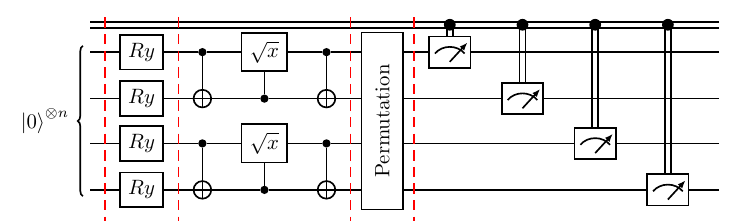} &
        \includegraphics[width=0.5\textwidth]{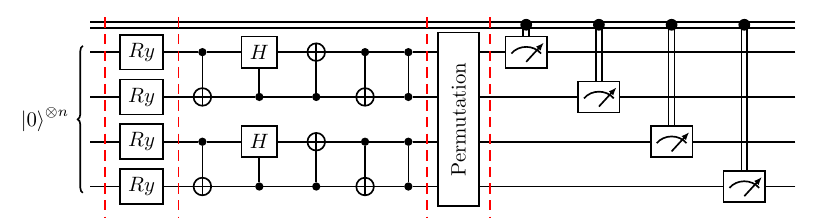} 
    \end{tabular}
    \caption{
        QLBM with quantum streaming solving (a) the diffusion equation and (b) Burgers' equation. 
        The permutation block encapsulates a sequence of SWAP gates that shift the qubits during the streaming step.
    }%
    \label{fig:qlga-b}%
\end{figure}

The first revision to the QLGA algorithm performs the streaming step using quantum operations, not through classical post-processing.
It can be implemented before the measurement step via the permutation function that applies a sequence of SWAP gates to shift the qubits to their neighbors. 
The corresponding quantum circuit is shown in \cref{fig:qlga-b} for the diffusion and Burgers' equations.

We validate the proposed variant by solving the 1D diffusion and Burgers' equations using Gaussian and sinusoidal distributions as the initial condition, respectively.
Periodic boundary conditions are used.
We compare the two QLBM solutions, i.e., with classical streaming and quantum streaming, with the classical solution using the D1Q2 scheme and $26$ simulated qubits. 
The simulations were performed using Qiskit AerSimulator with the statevector simulation method, and the counts were sampled from the statevector using $10^5$ shots. The spatial domain consists of $N_s=13$ lattice sites corresponding to $26$ qubits. Results are shown after $10$ time steps.

\begin{figure}[h]
    \centering
    \begin{tabular}{c c} 
        \multicolumn{1}{l}{\footnotesize (a)} & 
        \multicolumn{1}{l}{\footnotesize (b)} \\
        \includegraphics{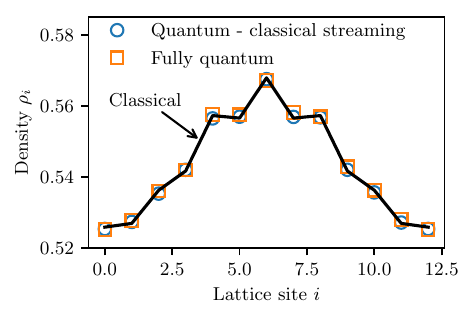} &
        \includegraphics{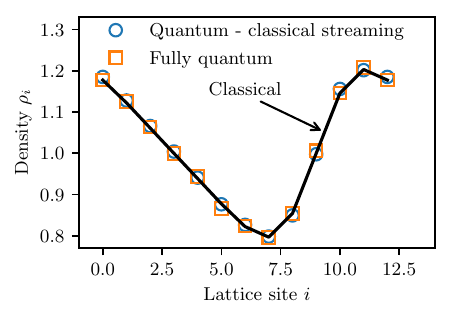}
    \end{tabular}
    \caption{
        QLBM solutions to (a) the diffusion equation and (b) Burgers' equation using the D1Q2 scheme.
        ``Quantum - classical streaming'' corresponds to the classical streaming algorithm discussed in previous work~\citep{yepezQuantumLatticegasModel2001}, ``Fully quantum'' refers to the present algorithm using direct measurements, and ``Classical'' is the classical implementation of the lattice method.
        The small differences between ``Quantum - classical streaming" and ``Fully quantum" are due to finite sampling in quantum measurement.
    }
    \label{fig:qlga-results-q2}
\end{figure}

The results of \cref{fig:qlga-results-q2} demonstrate good agreement with the classical solution. 
This variant removes the need for classical communication between lattice sites and does not cause global entanglement, as SWAP gates merely interchange the qubit quantum states. 
Since the quantum streaming does not change the local nature of product states, the total number of measurements required in each time step still scales linearly with the number of lattice sites.

In terms of complexity, we examine the relationship between the circuit depth and the system size by analyzing the different operations of the algorithm separately.
The encoding step consists of applying one rotation gate simultaneously to each qubit. Thus, changing the grid resolution and the qubit number does not affect the circuit depth. 
The same unitary operator is applied to each pair of qubits for the collision step. 
Thus, the circuit depth scales as $\mathcal{O}(1)$ with the number of qubits $n$.
For the streaming step implemented using the permutation function, the number of SWAP gates applied to one of the qubits increases linearly with the number of qubits. 
Therefore, the complexity of the algorithm is given by $\mathcal{O}(n)$. A D1Q3 scheme can also be implemented by using an extra qubit to create a smoother curve (\cref{fig:qlga-results-q3}), as shown in the \cref{s:d1q3}. 
For a D1Q2 problem with a point source, particles do not remain in the original lattice site but propagate through the lattices over time.
Therefore, a D1Q3 lattice scheme produces a smoother curve due to the non-zero distribution at one lattice site.

\subsection{Restless QLBM}

An issue with QLGA is that the quantum state collapses due to measurements performed during each time step, requiring re-encoding. 
However, we should allow the qubits to explore more complex states to achieve truly quantum computation. 
Furthermore, measurement on superconducting quantum computers can typically take an order of magnitude longer than gate times, adding significant time to the algorithm~\citep{QuantumAI}.
Thus, the goal is to eliminate repeated initialization of the quantum state and repeated measurements.

We attempted a unitary transformation using random rotations to reset the qubit relative phases after each time step.
However, this random phase kicking leads to dephasing, where the expected value of the off-diagonal elements of the density matrix decay to zero with time~\citep{nielsenQuantumComputationQuantum2010}. 
Therefore, a different approach is needed. 

To this end, our algorithm adds a phase-correction step, approximating the qubit relative phases and subtracting them at the end of each time step, delaying the need for measurement until the end of the computation. 
Phase correction is achieved via quantum phase estimation (QPE) as a sub-routine to estimate the eigen-phases of the encoding and collision unitaries~\citep{nielsenQuantumComputationQuantum2010}. 
These eigen-phases are then used to classically estimate the qubit relative phases, subtracted at the end of each time step.
The accuracy of the QPE depends on the number of ancillary qubits it uses. 
Thus, we use the iterative phase estimation (IPE) algorithm, which only requires a single auxiliary qubit and evaluates the phase bit by bit through a repetitive process~\citep{dobvsivcek2007arbitrary}. 
The accuracy of the IPE algorithm is determined by the number of iterations instead of the number of ancillary qubits. Therefore, they are of great importance for near-term quantum computing.
The IPE algorithm still requires measurement to estimate the phase. 
However, only one qubit, i.e., the auxiliary qubit, is measured, and not all qubits are in the quantum register.

Although the phase estimation algorithm could help avoid measurements at the end of each time step and re-preparation in the next time step, it introduces extra $O(n)$ mid-circuit measurements in each time step.
This overhead balances off the save for measurement.
What's left is the resource reduction for state re-preparation with $O(T)$ circuit depth, where $T$ is the number of time steps.

\begin{figure}
    \centering
    \begin{tabular}{c c} 
        \includegraphics[width=0.6\textwidth]{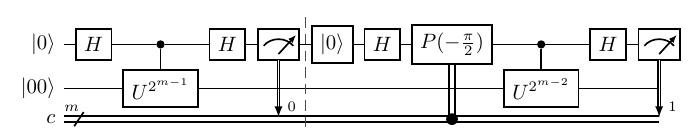} &
        \includegraphics[width=0.4\textwidth, height=3cm, keepaspectratio]{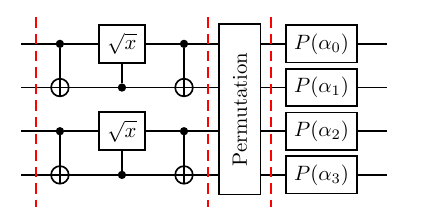} \\
        \multicolumn{1}{l}{\footnotesize (a)} & 
        \multicolumn{1}{l}{\footnotesize (b)} 
    \end{tabular}
    \caption{(a) IPE circuit used to estimate the qubits relative phase $\alpha_i$ (first two iterations) and (b) Restless QLBM circuit solving the diffusion equation.
    }%
    \label{fig:IPE}%
\end{figure}

In the context of QLBM, we apply the IPE algorithm to every lattice site, i.e., to a two-qubit unitary for the D1Q2 scheme. 
We set the unitary $U$ to be the product of the encoding and the collision operations since the streaming step only swaps the qubits and does not change their phases. 
For a total number of iterations $m$, the IPE algorithm estimates the best $m$ bit approximation to $\phi$ with $\varphi = 0.\varphi_1 ... \varphi_m = \sum_{k=1}^{m}\varphi_k/2^k$, where $\varphi_k$ are the phase bits ($1 \leq k \leq m$) in binary expansion.
At each iteration, $k$, the IPE algorithm applies the unitary $U$ $2^{m-k}$ times controlled on the auxiliary qubit.
Due to phase kickback, the relative phase of the auxiliary qubit becomes $\exp(i2\pi0.\varphi_{m-k+1}\varphi_{m-k+2}...\varphi_{m})$. 
Thus, to estimate the phase bit $\varphi_{m-k+1}$, a phase correction of $-2\pi(\varphi_{m-k+2}/2^2+...+\varphi_{m}/2^m)$ is performed, controlled on the measurement outcome of the previously estimated phase bits $\varphi_{m-k+2},\dots,\varphi_{m}$.
The phase bit $\varphi_{m-k+1}$ is then estimated by measuring the auxiliary qubit in the $x$-basis. 
The first two iterations of the IPE algorithm, which estimates the two least significant bits of the phase $\varphi_{m}$ and $\varphi_{m-1}$, is illustrated in~\cref{fig:IPE}~(a). 
One time step of the restless QLBM circuit is shown in the circuit of \cref{fig:IPE}~(b), which consists of collision, streaming, and phase correction operations. 

\begin{figure}[t]
    \centering
    \begin{tabular}{c c} 
        \includegraphics[width=0.43\textwidth]{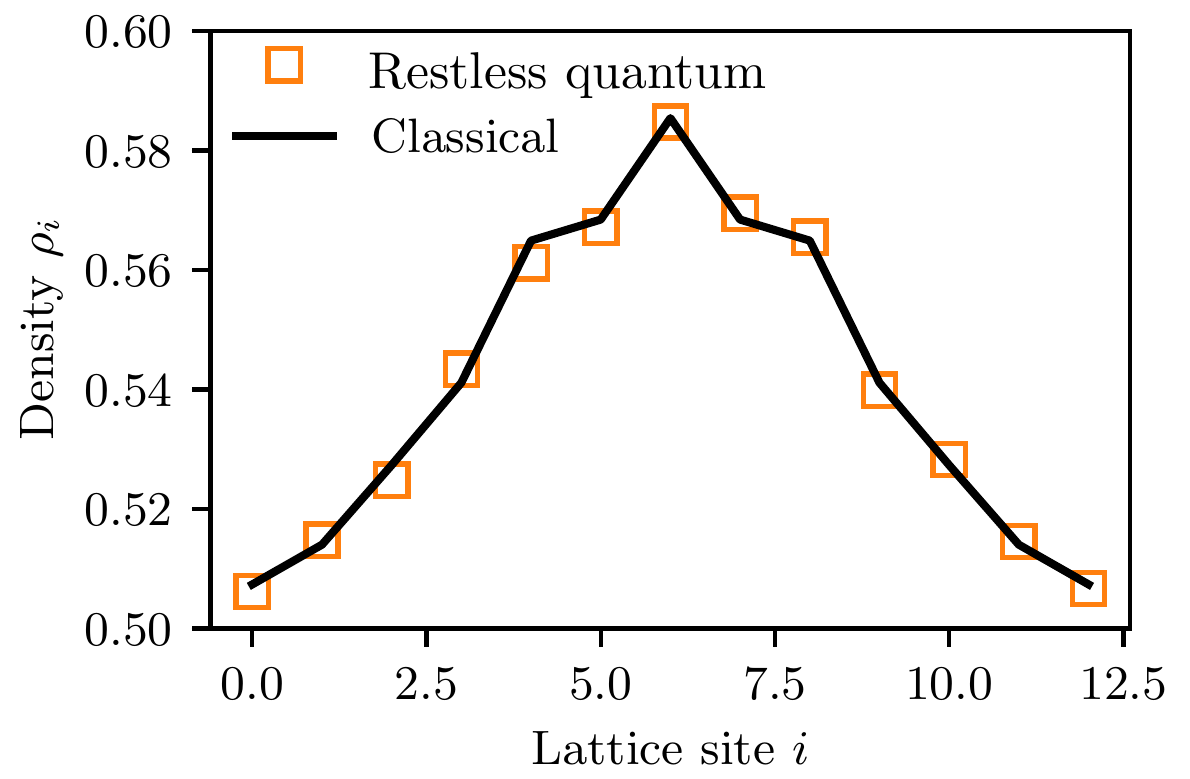} &
        \includegraphics[width=0.46\textwidth, keepaspectratio]{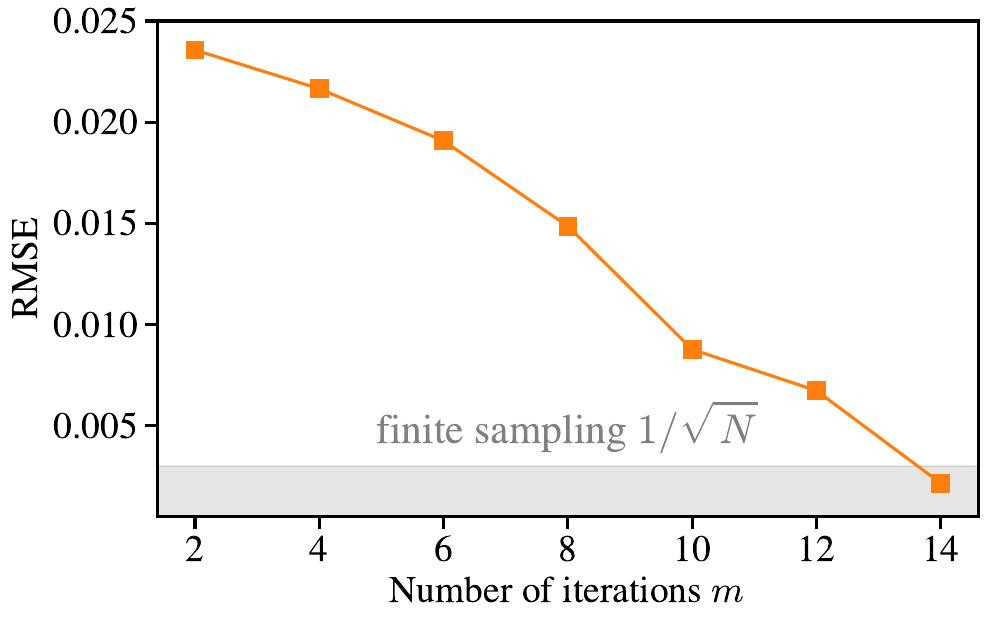} \\
        \multicolumn{1}{l}{\footnotesize (a)} & 
        \multicolumn{1}{l}{\footnotesize (b)} 
    \end{tabular}
    \caption{(a) Restless QLBM (present work) solution to the diffusion equation using the IPE algorithm with $m=14$ iterations, $N=10^5$ shots, and its comparison to classical results for the same method and resolution.
    (b) Comparing different number of iterations $m$ impact the overall algorithmic error.}
    \label{fig:qlga-results-q4}
\end{figure}

\Cref{fig:qlga-results-q4} presents the numerical simulation of our variant using the IPE algorithm to solve the 1D diffusion equation. 
The simulation uses the D1Q2 scheme with 26 qubits and runs on the statevector simulator. 
The spatial domain consists of $N_s=13$ lattice sites, and results are shown after $6$ time steps.
The quantum solution and the classical one agree well, with the root-mean-square error between them below the finite sampling threshold.
This variant improves the efficiency of the QLGA algorithm as the encoding is used only once at the beginning of the computation to prepare the quantum states, and measurement is only performed at the end of the computation. 

The number of iterations can be determined by a desired accuracy $1/2^q$ with probability of success at least $p$ 
\begin{gather} \label{eq:itr-ipe}
    m = q + \bigg\lceil \log_2\bigg(2+\frac{1}{2(1-p)}\bigg) \bigg\rceil.
\end{gather}
The difference between the estimated phase $\varphi$ and the actual phase $\phi$ satisfies $0\leq\Delta \varphi \leq 1/2^q$, where $\Delta \varphi = \phi -\varphi$.
The probability of success $p$ quantifies the likelihood one measures the best $m$ bit approximation $\varphi$~\citep{nielsenQuantumComputationQuantum2010}.
Such a probability is lower bounded by $4/\pi^2\approx 0.405$ but can be problematic in larger qubit cases.
Since the IPE algorithm only operates on each lattice site, running locally on each site requires $b+1$ qubits.
Hence, the desired accuracy $1/2^q$ is a more important factor. 
We numerically verify that it suffices to choose $q=12$ and $p=0.75$, hence $m=14$, to conduct a simulation comparable to classical solutions in \cref{fig:qlga-results-q4} (a), where the finite approximation error $N_s/2^q$ for phase estimation is smaller than the finite sampling threshold $1/\sqrt{N}$.
We further compare how different numbers of iterations $m$ impact the overall algorithmic error in \cref{fig:qlga-results-q4} (b).
For larger simulation, $q$ can be determined as $q\geq \log_2 (N_s\sqrt{N})$.
One can readily generalize this algorithm to a D$a$Q$b$ scheme, where the size of the quantum circuit for the IPE algorithm increases to $b+1$ qubits.
This change increases circuit complexity, though not necessarily the number of bits.

Regarding complexity, having only one ancillary qubit comes with the cost of $O(2^{m-1})$ gates, where $m$ is the number of IPE iterations related to the precision of the estimated phases. 
Given unbiased assumptions, the finite sampling error of iterative phase estimation scales proportional to $1/\sqrt{N}$, where $N$ is the shot number in quantum measurement.
A qubit state is multi-parametric, so subtracting the phase with a single estimation is more efficient.

\section{Results}\label{Sec5}

We test different quantum simulator implementations of our QLBM algorithm and its circuit for verification.
An in-depth discussion of the scaling of these simulators can be found in \cref{s:simres}.
In this section, we take measurements from each simulator to show the general trends of RMSE with respect to the quantum hardware constraints.

\begin{figure}[h]
    \centering
    \begin{tabular}{ c c }
        \includegraphics[]{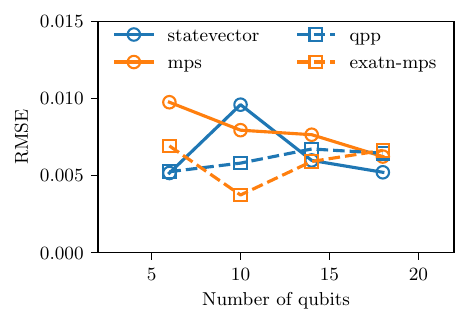} &
        \includegraphics[]{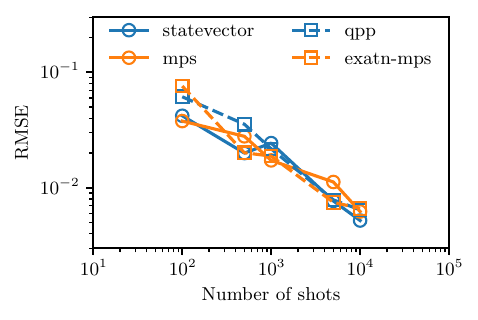} \\
        \multicolumn{1}{l}{\footnotesize (a)} & 
        \multicolumn{1}{l}{\footnotesize (b)} 
    \end{tabular}
    \caption{
        RMSE of the quantum solution solving the diffusion equation compared to the classical one. 
        (a) $10^4$~shots and (b) 18 qubits.
        Qiskit (solid lines, ------) statevector and MPS simulators and XACC  simulators (dashed lines, --~--~--) qpp and exatn-mps are shown.
        The colors correspond to (blue) statevector-like and (orange) matrix product state simulators.
        Notably, each lattice site contains $2$ qubits for a D1Q2 scheme.
    }%
    \label{fig:simulators}%
\end{figure}

We run the QLBM circuit of~\cref{fig:qlga-b} on the four different simulators and compare their performance. 
\Cref{fig:simulators} shows and the root-mean-squared error (RMSE),
\begin{gather}
    \mathrm{RMSE} = 
    \sqrt{
        \frac{
            \sum_{i = 0}^{N_s-1} \left(
                \rho_i^{\mathrm{(quant.)}} - 
                \rho_i^{\mathrm{(classic.)}}
            \right)^2
        }{N_s}
    },
\end{gather}
where $N_s$ is the number of lattice sites of the quantum solution compared to the classical one as a function of the number of shots (samples from the underlying density functions) and qubits.
The errors are reported from simulations as a function of the number of lattice sites in~\cref{fig:simulators}~(a) and as a function of the number of shots in~\cref{fig:simulators}~(b).

These figures demonstrate the scalability of this method.
The RMSE for the algorithm follows the trend of decreasing quadratically with the number of shots, as shown in \cref{fig:simulators}~(b).
An RMSE of $0.01$ occurs at less than $10^4$ shots.
This accuracy level is proportional to $1/\sqrt{N}$ due to finite sampling of the quantum state, where $N$ is the shot number.
The number of qubits does not affect the RMSE greatly, as shown in \cref{fig:simulators}~(a), and the technique can scale well to more lattice sites with a linear increase in qubits. 

\begin{figure}
    \includegraphics[]{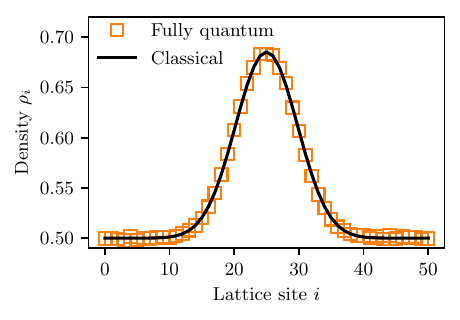}
    \caption{QLBM solution using D1Q3 scheme to the diffusion equation using the Qiskit MPS simulator and its comparison to classical results using $102$ qubits.
    }
    \label{fig:diffusion_102qubits_mps}
\end{figure}

The MPS simulator uses a local representation in a factorized form of tensor products~\citep{vidal2003efficient}. 
MPS ensures that the overall structure remains small as long as the circuit has a low degree of entanglement, enabling more simulated qubits.
To demonstrate this, we run the QLBM circuit of~\cref{fig:qlga-b}~(a) on the Qiskit MPS simulator for a resolution of $51$ lattice sites corresponding to $102$ qubits using $10^6$ shots.
Results after $10$ time steps are shown in \cref{fig:diffusion_102qubits_mps} and demonstrate good agreement with the classical solution. 

\section{Conclusion and Discussion}\label{s:conclusion}

In this work, we present a revised quantum lattice gas algorithm that can be used to simulate fluid flows.
The algorithm of \citet{yepezQuantumComputationFluid} was designed for Type-II quantum computers that are no longer used.
The presented algorithm uses a full-quantum implementation of the streaming step to reduce classical communication between lattice sites.
In addition, quantum streaming optimizes repetitive encoding operations by estimating the qubit relative phases and subtracting them at the end of each time step.
State re-preparation and measurements can scale exponentially in modern computers, and our work on state preparation optimization is essential for efficiently adapting the QLGA to modern hardware.
The algorithm is tested to solve canonical PDEs: the diffusion and Burgers' equations.
We explored different quantum simulators available on Qiskit and XACC quantum computing frameworks and compared their performance and wall-clock time.
Matrix product state simulators are most efficient for the presented quantum circuits, as they have low and local entanglement.
These strategic algorithm improvements enable the execution of the quantum lattice gas algorithm on quantum hardware, a subject of future work.

An important limitation of the current algorithm is the size of the required quantum register, which scales linearly with the number of lattice sites. 
For example, the number of required qubits in a two-dimensional $M\times M$ lattice is $5M^2$ using the D2Q5 scheme. 
This requirement constrains the ability to extend this algorithm to higher dimensions due to the limited number of qubits available on current quantum hardware. 
Thus, more appropriate encoding schemes, such as amplitude-based encoding, are needed to scale logarithmically with the number of lattice sites.
For example, \citet{budinskiQuantumAlgorithmAdvection2021} implements the streaming step via a quantum random walk, similar to the collisionless Boltzmann equation solved on a quantum computer of \citet{todorova2020quantum}.
Indeed, in principle, one could design an algorithm that uses a quantum random walk for both the collision and streaming operations, which we will investigate in future work.  

\section*{Acknowledgements}

We thank Dr.~Jeffrey Yepez for numerous fruitful discussions of this work.
SHB acknowledges support from the Georgia Tech Quantum Alliance and a Georgia Tech Seed Grant.
EFD acknowledges DOE ASCR funding under the Quantum Computing Application Teams program, FWP number ERKJ347.
This work used Bridges2 at the Pittsburgh Supercomputing Center (PSC) and Delta and the National Center for Supercomputing Applications (NCSA) through allocation PHY210084 from the Advanced Cyberinfrastructure Coordination Ecosystem: Services \& Support (ACCESS) program, which is supported by National Science Foundation grants \#2138259, \#2138286, \#2138307, \#2137603, and \#2138296.
This research used resources of the Oak Ridge Leadership Computing Facility, which is a DOE Office of Science User Facility supported under Contract DE-AC05-00OR22725.

\section*{Data availability}

The data and code supporting this study's findings are available at \url{https://github.com/comp-physics/fully-QLBM} under the MIT license.

\section*{Conflict of interest}

The authors have no conflicts to disclose.

\bibliography{main.bib}

\appendix

\section{D1Q3 Implementation}\label{s:d1q3}
\begin{figure}[h]
    \includegraphics[]{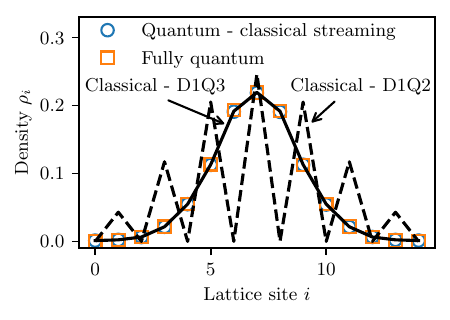}
    \caption{\label{fig:qlga-results-q3} QLBM solutions to the diffusion equation using the D1Q3 scheme.
    ``Quantum - classical streaming'' corresponds to the classical streaming algorithm discussed in previous work~\citep{yepezQuantumLatticegasModel2001}, ``Fully quantum'' refers to the present algorithm, and ``Classical'' is the classical implementation of the lattice methods as labeled.
    }
\end{figure}

The implementations of the QLBM mentioned in \cref{Sec2} and \cref{Sec3} used the D1Q2 scheme, which uses two moving particles per lattice site. 
This scheme shows checkerboard pathologies when using a sharp function as the initial condition because it simulates two independent sub-lattice~\citep{yepezQuantumLatticegasModel2001}.
\citet{yepezQuantumLatticegasModel2001a} presents a cure for this, allowing only one qubit to move while the other remains stationary.
However, this requires two collision and streaming procedures per time step. 
Our results show that the quantum D1Q3 scheme, which adds a stationary qubit to each lattice site, solves this problem.
A D1Q3 scheme has three particle distributions across 3 links: left, right, and stationary.
So, some fictitious particles remain stationary in a time-step, smoothing the curve.
\Cref{fig:qlga-results-q3} compares the quantum solution of the diffusion equation using the D1Q2 scheme with the D1Q3 scheme for a delta function as the initial condition.  
These simulations were performed using the Qiskit matrix product state simulator, $45$ qubits for the D1Q3 scheme, and the sampling procedure used $10^6$ shots. 
The spatial domain consists of $N_s=15$ lattice sites, and results are shown after $10$ time steps.
The result shows that the quantum D1Q3 solution agrees with the classical one and that the D1Q3 version remedies the checkerboard effect. 
We note that the D1Q3 scheme does not affect algorithmic complexity and can employ measurement avoidance analogous to the D1Q2 scheme presented in the main manuscript body.

\section{Simulator Results}\label{s:simres}

\begin{figure}[h]
    \includegraphics[]{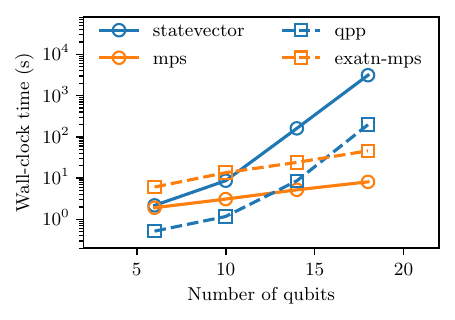}
    \caption{\label{fig:qlga-results-wall} Wall-clock time in seconds using $10^4$~shots.
    }
\end{figure}

In \cref{fig:qlga-results-wall}, we show the results for different quantum simulator implementations of our algorithm and its circuit.
Simulators include statevector-based and tensor-network types that use the matrix product state (MPS) method~\citep{perez2006matrix}.
Two different simulation frameworks are tested: Qiskit and XACC~\citep{mccaskey2017extreme}.
Statevector and MPS simulators are part of the Qiskit Aer backend. 
The XACC qpp (Quantum++) simulator is based on a C++ general-purpose quantum computing library~\citep{gheorghiu2018quantum}.
XACC exatn-mps simulator is a part of the TNQVM (tensor-network quantum virtual machine) simulation backend and uses a noiseless, matrix product state (MPS) wave function decomposition for the quantum circuit~\citep{mccaskey2018validating}. 
The wall-clock times of \cref{fig:simulators}~(c) were obtained via simulation on an AMD~EPYC~7742 (64-core) processor.

We find that the results are shot-noise limited, and all quantum simulators obtain the same accuracy in terms of RMSE, which decreases from less than $10\%$ for $10^2$ shots to less than $1\%$ for $10^4$ shots.
However, the Qiskit MPS simulator is at least one order of magnitude faster than the statevector ones.
This behavior is expected.
The Qiskit statevector and XACC (qpp) simulators generate the full state vector, which scales exponentially with the number of qubits.

\end{document}